\documentclass{aastex702}
\usepackage{xurl}
\hypersetup{hidelinks}

\shorttitle{Electron VDFs at Low-$\beta$ Quasi-Perpendicular Shocks}

\accepted[]{This article has been accepted for publication in The Astrophysical Journal.}

\begin{document}

\title{Formation of Electron Velocity Distribution Functions at Low-$\beta$ Quasi-Perpendicular Shocks From Physical-Mass-Ratio PIC Simulations}

\author{Zhoufan Mu}
\affiliation{Institute of Frontier and Interdisciplinary Science, Shandong University, Shandong, China}
\email[show]{202217045@mail.sdu.edu.cn}

\author{Yao Chen}
\affiliation{Institute of Frontier and Interdisciplinary Science, Shandong University, Shandong, China}
\affiliation{Shandong Key Laboratory of Space Environment and Exploration Technology, Shandong University, Shandong, China}
\affiliation{Institute of Space Sciences, Shandong University, Shandong, China}
\email[show]{yaochen@sdu.edu.cn}
\correspondingauthor{Yao Chen}

\author{Hao Ning}
\affiliation{Institute of Frontier and Interdisciplinary Science, Shandong University, Shandong, China}
\email[show]{haoning@sdu.edu.cn}

\author{Shan Wang}
\affiliation{Peking University, Beijing, China}
\email[show]{shanwang90@pku.edu.cn}

\author{Sibo Xu}
\affiliation{School of Earth and Space Sciences, Peking University, Beijing, China}
\email[show]{xusibo@pku.edu.cn}

\author{Terry Zixu Liu}
\affiliation{Shandong Key Laboratory of Space Environment and Exploration Technology, Shandong University, Shandong, China}
\affiliation{Institute of Space Sciences, Shandong University, Shandong, China}
\email[show]{terryliuzixu@sdu.edu.cn}

\author{Ao Guo}
\affiliation{University of Science and Technology of China, Hefei, Anhui, China}
\email[show]{Aoguo19@mail.ustc.edu.cn}

\author{Hui Zhang}
\affiliation{School of Space Science and Technology, Shandong University, Weihai, Shandong, China}
\email[show]{h.zhang@sdu.edu.cn}

\author{Hui Zhu}
\affiliation{Institute of Frontier and Interdisciplinary Science, Shandong University, Shandong, China}
\email[show]{huizhu@email.sdu.edu.cn}

\begin{abstract}
At low-$\beta$ quasi-perpendicular shocks, the origins of upstream field-aligned electron beams and downstream flat-top distributions remain unresolved. Using 2D fully kinetic particle-in-cell (PIC) simulations at the physical ion-to-electron mass ratio ($m_{\mathrm{i}}/m_{\mathrm{e}} = 1836$), we found that the shock transition region is dominated by electromagnetic whistler-mode turbulence driven by the modified two-stream instability (MTSI) of reflected ions, not by the electrostatic instabilities as suggested by some previous studies using a reduced mass ratio. Beams in the upstream are produced mainly by shock-drift acceleration (SDA) at the ramp, whereas we attribute the downstream flat-top distributions to nonuniform acceleration by the parallel electric field. Electrons experience different field strengths, polarities, and interaction durations along their trajectories, leading to a spread of parallel velocity changes that broadens and flattens the distribution. These results highlight the importance of using the physical mass ratio in exploring the electron dynamics of low-$\beta$ shocks.
\end{abstract}

\section{Introduction}

Quasi-perpendicular collisionless shocks are major sites of electron acceleration in the heliosphere. Field-aligned electron beams are commonly observed upstream of Earth's bow shock \citep{fredricks1968,fredricks1970,feldman1982,feldman1983a,bale1998,chen2018} and at interplanetary shocks driven by coronal mass ejections (CMEs) \citep{gurnett1979,simnett2003}. Magnetospheric Multiscale (MMS) observations at Earth's bow shock suggest that field-aligned beams can also act as a seed population, which is then accelerated in a shrinking magnetic bottle \citep{lindberg2023}. These beams drive wave-particle interactions in the upstream region. They can excite Langmuir turbulence and density fluctuations, whose nonlinear conversion produces plasma emission, including type II radio bursts that provide remote-sensing diagnostics of shock dynamics \citep{wild1950,melrose1980,melrose2017,cane1981,holman1983,bale1999}. MMS measurements in the electron foreshock indicate that, besides three-wave decay, small-scale density fluctuations in the ambient plasma are important for the evolution of Langmuir waves \citep{graham2026}.

Shock-drift acceleration (SDA), also known as fast Fermi acceleration, is the classical mechanism proposed to generate the upstream field-aligned beams \citep{wu1984}. In the normal incidence frame, the magnetic field gradient drives an electron drift along the shock surface, and the motional electric field does work on this drift. In the de Hoffmann--Teller (HT) frame, the same reflection appears as adiabatic mirroring, with kinetic energy conserved when the cross-shock potential and wave fields are neglected \citep{kraussvarban1989,amano2022}. Reflection produces a characteristic parallel speed of order $2v_{\mathrm{sh}}/\cos\theta_{Bn}$ in the upstream rest frame, where $v_{\mathrm{sh}}$ is the shock speed relative to the upstream plasma. Under low-$\beta$ conditions, the lower electron thermal speed reduces the fraction of electrons that satisfy the reflection condition. In the transition region, the turbulence becomes stronger, leading to stronger electron scattering and a shorter effective acceleration time \citep{armstrong1977,holman1983}. These combined effects challenge the role of SDA in beam formation.

Recent kinetic particle-in-cell (PIC) simulations reported various scenarios for beam formation at low-$\beta$ quasi-perpendicular shocks. \citet{guo2024} showed that relative drift between reflected ions and incident electrons excites the electron cyclotron drift instability (ECDI) at the leading edge of the shock foot; the ECDI waves enhance SDA by scattering electrons out of the loss cone and strengthen the upstream field-aligned beam. \citet{guo2025} proposed that electron shear flow at the shock ramp drives the electron Kelvin-Helmholtz instability (EKHI), forming electron vortices. Charge separation within these vortices generates a parallel electric field that directly accelerates electrons into field-aligned beams.

{\sloppy Both studies used a reduced ion-to-electron mass ratio of $m_{\mathrm{i}}/m_{\mathrm{e}} = 100$, which can qualitatively alter the dominant instability. The key parameter is the ratio of the reflected-ion drift speed $v_{\mathrm{d}}$ to the electron thermal speed $v_{\mathrm{th,e}}$, $v_{\mathrm{d}}/v_{\mathrm{th,e}} \approx M_{\mathrm{A}}/\sqrt{\beta_{\mathrm{e}} \cdot m_{\mathrm{i}}/m_{\mathrm{e}}}$ \citep{umeda2012}. At $m_{\mathrm{i}}/m_{\mathrm{e}} = 100$, $v_{\mathrm{d}}/v_{\mathrm{th,e}} \sim 4$-5, favoring electrostatic beam instabilities such as the ECDI. At the physical mass ratio, $v_{\mathrm{d}}/v_{\mathrm{th,e}} \sim 1$, where electrostatic modes are suppressed and the electromagnetic MTSI becomes dominant \citep{scholer2004,umeda2012}. This transition is consistent with full-particle simulations, which find electromagnetic instabilities for $m_{\mathrm{i}}/m_{\mathrm{e}} \gtrsim 625$ \citep{scholer2003,scholer2004,scholer2006,umeda2012,umeda2018,xu2026}. Whether SDA can still generate upstream electron beams when the instability regime changes at the physical mass ratio remains unknown.\par}

Flat-top electron velocity distributions are commonly observed downstream of quasi-perpendicular shocks \citep{montgomery1970,feldman1982,feldman1983a,feldman1983b,fitzenreiter2003}. They are characterized by an approximately uniform phase-space density over a broad range of parallel velocities, with greater mean energy in the parallel than in the perpendicular direction. Several mechanisms have been proposed, including direct-current (DC) field energization inferred from test-particle trajectories in prescribed shock structures \citep{scudder1986,thomsen1987,scudder1995}, stochastic wave energization (SWE) inferred from spacecraft observations at Earth's bow shock \citep{stasiewicz2023}, and wave-particle trapping by electrostatic waves in one-dimensional unmagnetized Vlasov simulations \citep{oneil1965,zanelli2025}. How flat-top distributions form in the electromagnetic turbulence of fully kinetic shocks at the physical mass ratio remains unresolved.

Here we use two-dimensional fully kinetic PIC simulations with the physical ion-to-electron mass ratio ($m_{\mathrm{i}}/m_{\mathrm{e}} = 1836$) and upstream parameters chosen to facilitate comparison with \citet{guo2025}. We investigate upstream beam and downstream flat-top formation in the resulting electromagnetic turbulence. We combine shock diagnostics, wave analysis, and particle tracing to examine how SDA produces the upstream beam and how nonuniform acceleration by the parallel electric field forms the downstream flat-top distribution.

\section{Simulation Setup}

We performed two particle-in-cell (PIC) simulations with two spatial and three velocity dimensions (2D3V) in the $x$-$z$ plane using the open-source code vPIC \citep{bowers2008,bowers2009}, both with the physical ion-to-electron mass ratio ($m_{\mathrm{i}}/m_{\mathrm{e}} = 1836$). We designate the longer-duration, lower-resolution simulation as Case~L and the shorter-duration, higher-resolution simulation as Case~S. Case~L follows shock evolution over an extended interval, while Case~S resolves electron-scale turbulence and electron velocity distributions across the shock. Both cases used identical upstream conditions: $\beta_{\mathrm{e}} = \beta_{\mathrm{i}} = 0.01$, $\theta_{Bn} = 85^\circ$, $M_{\mathrm{A}} \approx 5$ (measured from shock front motion), $\omega_{\mathrm{pe}}/\Omega_{\mathrm{ce}} = 10$, and $v_{\mathrm{th,e}} = 0.01c$ ($T_{\mathrm{e}} \approx 0.3$ MK, corresponding to $k_{\mathrm{B}}T_{\mathrm{e}} \approx 26$ eV). The adopted values give a total $\beta$ of 0.02 and a proton $\beta$ of 0.01, which correspond to CME-driven shocks close to the Sun rather than to interplanetary shocks farther out; the simulations therefore describe electron VDF formation at a low-$\beta$ shock. The upstream magnetic field is $B_0 = (B_0\cos\theta_{Bn}\,\hat{z} + B_0\sin\theta_{Bn}\,\hat{y})$.

The upstream electron and ion inertial lengths are defined as $d_{\mathrm{e}}=c/\omega_{\mathrm{pe}}$ and $d_{\mathrm{i}}=c/\omega_{\mathrm{pi}}$, respectively. For the quasi-neutral electron--proton plasma considered here, $d_{\mathrm{i}}/d_{\mathrm{e}}=\sqrt{m_{\mathrm{i}}/m_{\mathrm{e}}}=\sqrt{1836}\approx42.85$.

Case~S has a domain of $L_x\times L_z=90\,d_{\mathrm{e}}\times270\,d_{\mathrm{e}}$ (approximately $2.10\,d_{\mathrm{i}}\times6.30\,d_{\mathrm{i}}$). Plasma is continuously injected from the right boundary ($z = L_z$) at $V_{\mathrm{in}} = 3V_{\mathrm{A}0}$, where $V_{\mathrm{A}0} = 0.0023c$, and reflected at the left boundary ($z = 0$), driving a shock toward $+z$. The $x$ boundaries are periodic. The grid spacing in both directions is $\Delta x=\Delta z=0.087\,d_{\mathrm{e}}\approx0.00203\,d_{\mathrm{i}}$, the time step is $\Delta t = 0.05\,\omega_{\mathrm{pe}}^{-1}$, and each cell initially contains 500 electrons and 500 ions, corresponding to approximately $3.2 \times 10^{9}$ macroparticles. Case~S was evolved to $t = 24000\,\omega_{\mathrm{pe}}^{-1}$.

Case~L has the same upstream conditions and boundary configuration as Case~S, with $L_x\times L_z=90\,d_{\mathrm{e}}\times720\,d_{\mathrm{e}}$ (approximately $2.10\,d_{\mathrm{i}}\times16.80\,d_{\mathrm{i}}$) and grid spacing $\Delta x=\Delta z=0.174\,d_{\mathrm{e}}\approx0.00406\,d_{\mathrm{i}}$ in both directions, and $\Delta t = 0.1\,\omega_{\mathrm{pe}}^{-1}$. Case~L was evolved to $t = 100000\,\omega_{\mathrm{pe}}^{-1}$ ($\approx 5.45\,\Omega_{\mathrm{ci}}^{-1}$) to follow the long-term shock evolution. The wave and particle analyses below use Case~S unless Case~L is explicitly specified. Throughout the following analysis, spatial scales are expressed in units of $d_{\mathrm{e}}$, wavenumbers as $kd_{\mathrm{e}}$, and wave frequencies as $\omega/\Omega_{\mathrm{e}}$, where $\Omega_{\mathrm{e}}\equiv\Omega_{\mathrm{ce}}$.

\section{Simulation Results}

\subsection{Shock Structure and Evolution}
\label{sec:shock_structure}

\begin{figure}[htbp]
\centering\includegraphics[width=\linewidth,height=0.68\textheight,keepaspectratio]{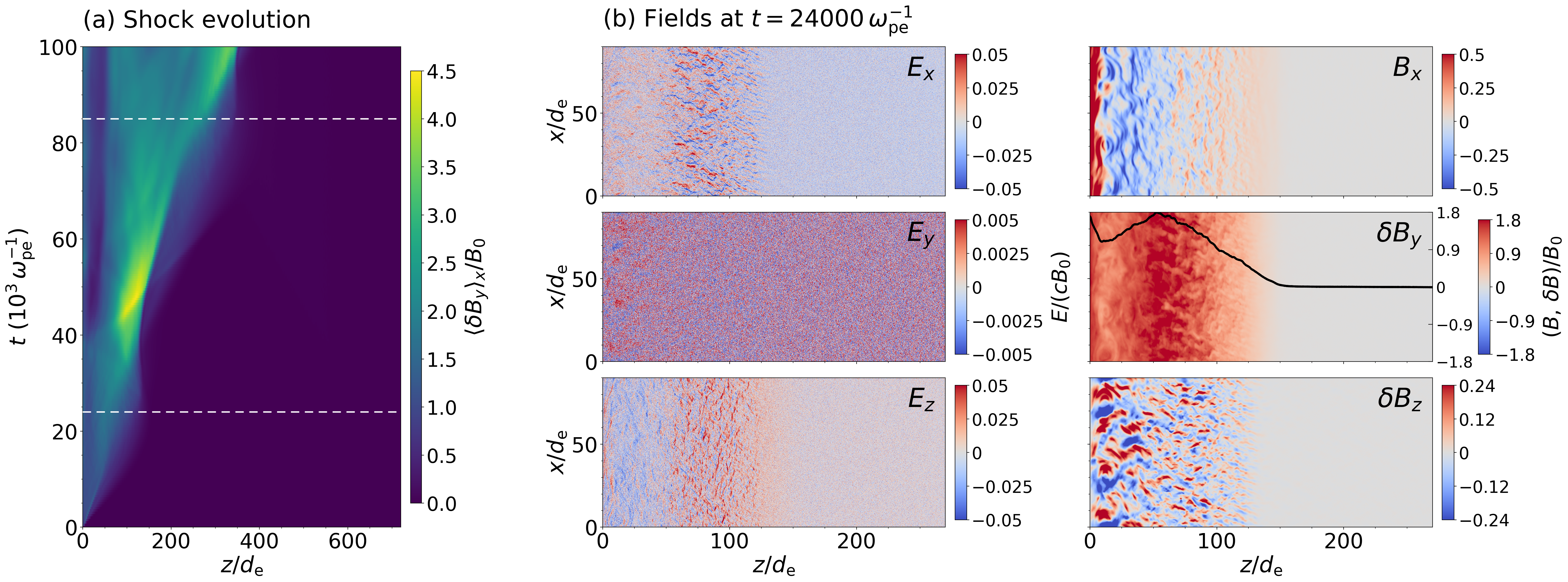}
\caption{Long-term shock evolution and electromagnetic field structure. (a) Space--time distribution of $\langle B_y(t)-B_y(0)\rangle_x/B_0$ from Case~L, averaged over the transverse direction $x$. The white dashed lines mark $t=24000\,\omega_{\mathrm{pe}}^{-1}$ and $t=85000\,\omega_{\mathrm{pe}}^{-1}$, corresponding to the snapshot times in panel (b) and Figure~\ref{fig:shockevolution}, respectively. (b) Two-dimensional electromagnetic fields from Case~S at $t=24000\,\omega_{\mathrm{pe}}^{-1}$ in the developed shock. Electric and magnetic fields are normalized to $cB_0$ and $B_0$, respectively. The initial field is subtracted from the $y$ and $z$ components of the magnetic field. The black curve shows $\langle\delta B_y\rangle_x/B_0$ as a function of $z$, where $\delta B_y=B_y(t)-B_y(0)$ and $\langle\cdots\rangle_x$ denotes averaging over the transverse direction $x$. Its amplitude is read from the right-hand vertical axis; the color maps are plotted against $z/d_{\mathrm{e}}$ and $x/d_{\mathrm{e}}$.}
\label{fig:fields}
\end{figure}

We first examine Case~L to place the snapshot from Case~S in the context of shock evolution. Figure~\ref{fig:fields}a shows the space--time distribution of $\langle B_y(t)-B_y(0)\rangle_x/B_0$, where the average is taken along the transverse direction $x$. As the shock propagates toward $+z$, the magnetic enhancement near the front strengthens, weakens, and subsequently strengthens again. This sequence indicates that Case~L encompasses at least one shock-reformation cycle. The white dashed lines mark $t=24000\,\omega_{\mathrm{pe}}^{-1}$ and $t=85000\,\omega_{\mathrm{pe}}^{-1}$, the times selected for the field comparisons below.

Figure~\ref{fig:fields}b presents the two-dimensional fields from Case~S at $t=24000\,\omega_{\mathrm{pe}}^{-1}$. At this time, the shock exhibits a developed foot--ramp--overshoot structure connecting the upstream and downstream plasmas. The black profile, obtained by averaging $\delta B_y=B_y(t)-B_y(0)$ along $x$, delineates the magnetic compression and its variation along the direction $z$ normal to the shock. Within the ramp, $E_z$ points upstream and reverses across the overshoot, forming the characteristic cross-shock potential. Electromagnetic fluctuations occupy both the transition region and the downstream plasma.

We also examined the shock structure at $t=85000\,\omega_{\mathrm{pe}}^{-1}$ in Case~L (Figure~\ref{fig:shockevolution}) and found its overall field morphology consistent with that in Case~S at $t=24000\,\omega_{\mathrm{pe}}^{-1}$ (Figure~\ref{fig:fields}b). This agreement supports using Case~S to examine the electron-scale structure of a developed shock.

\subsection{Electromagnetic Turbulence and Wave Properties}
\label{sec:turbulence}

\begin{figure}[htbp]
\centering\includegraphics[width=\linewidth,height=0.68\textheight,keepaspectratio]{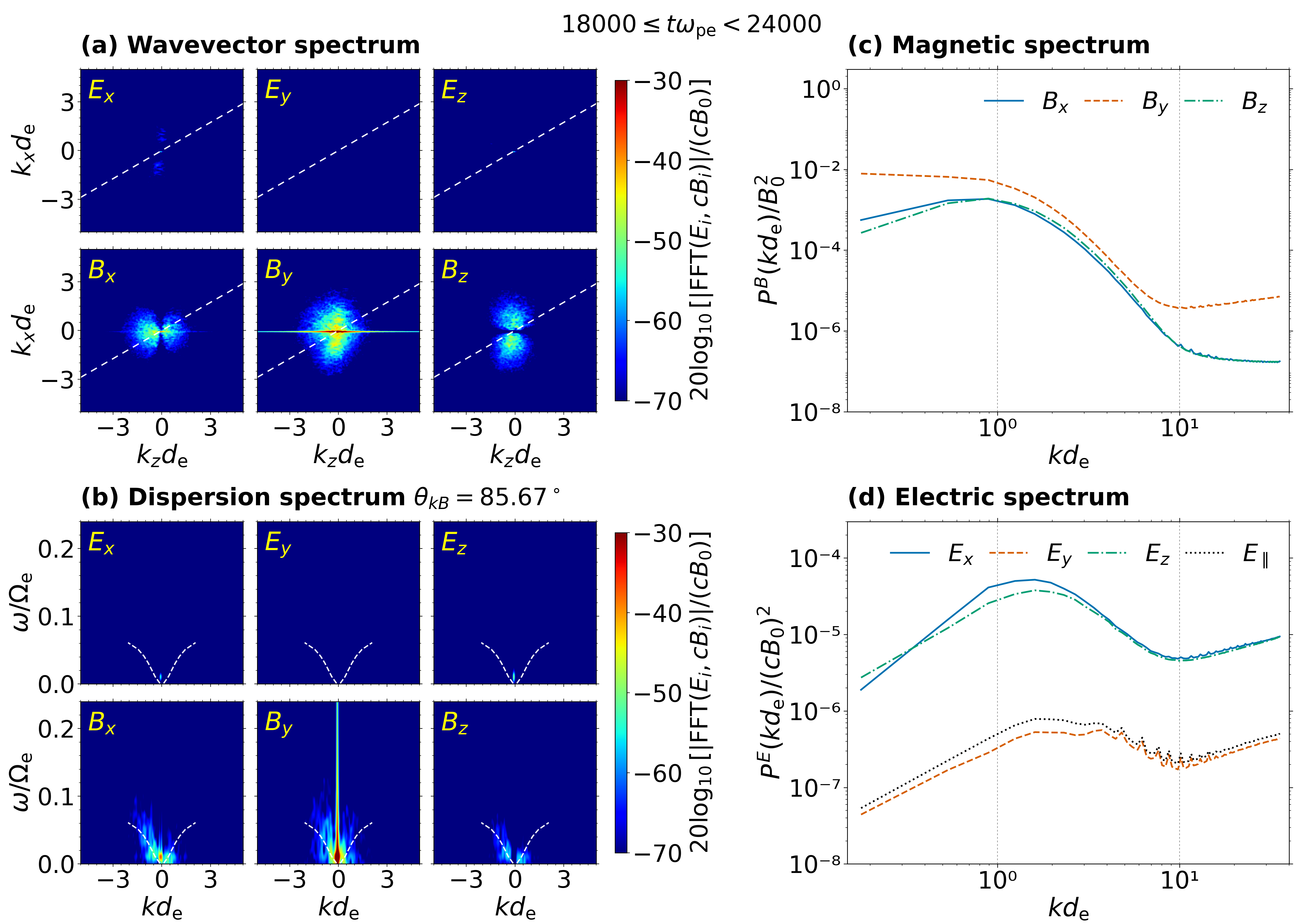}
\caption{Wave modes and spectra in the shock transition of Case~S ($18000 \leq t\omega_{\mathrm{pe}} < 24000$; $x$: $0$--$90\,d_{\mathrm{e}}$, $z$: $50\text{--}150\,d_{\mathrm{e}}$). (a) Frequency-maximized Fourier amplitudes in $k_x$--$k_z$ space, shown on the logarithmic color scale. White dashed lines mark the $30^\circ$ slice measured from the positive $k_z$ axis. (b) $\omega$--$k$ spectrum along this slice, corresponding to $\theta_{kB}=85.67^\circ$; white dashed curves show the cold-plasma whistler dispersion. Wavenumbers are expressed as $kd_{\mathrm{e}}$, and frequency as $\omega/\Omega_{\mathrm{e}}$, with $\Omega_{\mathrm{e}}=\Omega_{\mathrm{ce}}$. (c) Magnetic spectra of $B_x$, $B_y$, and $B_z$. (d) Electric spectra of $E_x$, $E_y$, and $E_z$, with $E_\parallel$ relative to $B_0$ shown as a black dotted curve. Panels (c)--(d) show radial shell-integrated power densities per unit $kd_{\mathrm{e}}$, normalized to $B_0^2$ and $(cB_0)^2$, respectively. Gray dashed vertical lines mark $1 \leq kd_{\mathrm{e}} \leq 10$. Fits of $P(kd_{\mathrm{e}})\propto(kd_{\mathrm{e}})^{-\alpha}$ over this interval give $\alpha_{Bx} = 4.25$, $\alpha_{By} = 3.72$, $\alpha_{Bz} = 4.34$, $\alpha_{Ex} = 1.41$, $\alpha_{Ey} = 0.57$, $\alpha_{Ez} = 1.28$, $\alpha_{E\parallel} = 0.72$.}
\label{fig:rampturbulence}
\end{figure}

In Case~S, broadband electromagnetic turbulence is present in both the shock transition and downstream regions over $18000\leq t\omega_{\mathrm{pe}}<24000$ (Figures~\ref{fig:rampturbulence} and \ref{fig:downstreamturbulence}). In both regions, the magnetic spectra follow power-law forms over $1\leq kd_{\mathrm{e}}\leq10$, with $B_y$ dominating over $B_x$ and $B_z$. The spectrum of the electric field parallel to $B_0$, $E_\parallel$, is flatter than the magnetic spectra over the same range.

In the shock transition region, the dispersion spectrum evaluated in the simulation laboratory frame follows the cold-plasma whistler branch over $1\leq kd_{\mathrm{e}}\leq3.5$ (Figure~\ref{fig:rampturbulence}b), supporting the identification of these fluctuations as whistler-mode waves. The $30^\circ$ slice measured from the positive $k_z$ axis corresponds to $\theta_{kB}\approx85.67^\circ$, indicating highly oblique propagation relative to the upstream magnetic field.

The polarization of the positive-$k$ branch further supports this identification. We integrate the magnetic spectral tensor over the same wavenumber range and within one frequency interval ($\Delta\omega/\Omega_{\mathrm{e}}\approx0.0105$) on either side of the whistler curve. Only 0.67\% of the magnetic power lies along $\mathbf{k}$. The magnetic compressibility, defined as the fraction of magnetic power along $\mathbf{B}_0$ in this band, is approximately 0.63. These properties are compatible: at highly oblique propagation, fluctuations transverse to $\mathbf{k}$ can have a substantial component along $\mathbf{B}_0$. The transverse components have a band-integrated phase difference of approximately $-91^\circ$. With the Fourier convention $\exp[i(\mathbf{k}\cdot\mathbf{x}-\omega t)]$ and the right-handed basis $\mathbf{e}_1\propto\hat{\mathbf{k}}\times\hat{\mathbf{b}}_0$, $\mathbf{e}_2=\hat{\mathbf{k}}\times\mathbf{e}_1$, the normalized magnetic helicity is $+0.27$, with the same rotation sense as the cold-plasma whistler solution. It remains positive (0.19--0.55) for band half-widths of 0.5--2 frequency intervals, indicating a preferred sense of rotation consistent with whistler waves within a spectrum of mixed polarization.

\subsection{Electron VDFs across the Shock}

\begin{figure}[htbp]
\centering\includegraphics[width=\linewidth,height=0.68\textheight,keepaspectratio]{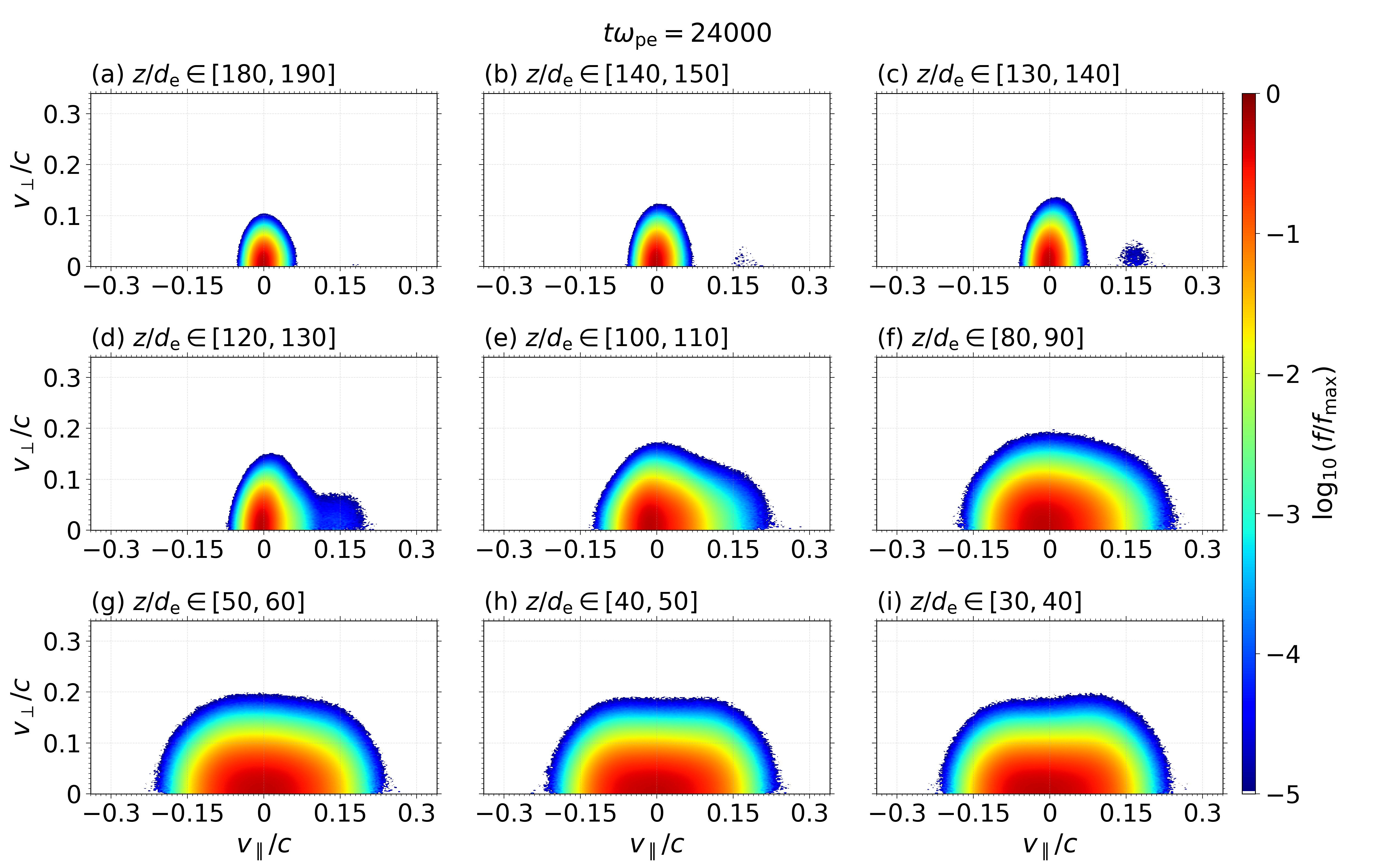}
\caption{Electron velocity distribution functions (VDFs) across the shock in Case~S at $t = 24000\,\omega_{\mathrm{pe}}^{-1}$, averaged over $10\,d_{\mathrm{e}}$ slices: (a-c) upstream; (d) transition; (e-f) toward the overshoot; (g-i) downstream. Velocities are normalized to $c$, with the mean parallel velocity subtracted in each slice. The decomposition is relative to $\mathbf{B}_0$, and $v_\perp$ is the magnitude of both perpendicular components.}
\label{fig:vdf}
\end{figure}

Figure~\ref{fig:vdf} shows electron VDFs from Case~S at $t\omega_{\mathrm{pe}}=24000$, sampled within slices of width $10\,d_{\mathrm{e}}$. The laboratory-frame velocities are resolved relative to the upstream magnetic field $\mathbf{B}_0$. The plotted parallel velocity is centered by subtracting its mean within each slice, while $v_\perp=\sqrt{v_{\perp1}^2+v_{\perp2}^2}\geq0$ denotes the magnitude of the two perpendicular velocity components. Here $v_{\perp1}$ and $v_{\perp2}$ are components along two mutually orthogonal directions perpendicular to $\mathbf{B}_0$. In the upstream region ($z\approx180\,d_{\mathrm{e}}$, approximately $30\,d_{\mathrm{e}}$ or $0.70\,d_{\mathrm{i}}$ ahead of the ramp), the electron distribution remains nearly Maxwellian (Figure~\ref{fig:vdf}a). Closer to the shock ($z$: $130$-$150\,d_{\mathrm{e}}$), a field-aligned beam emerges at $v_\parallel \approx 0.15$-$0.2c$ (Figures 3b-c). Across the transition region ($z$: $120$-$130\,d_{\mathrm{e}}$), the beam weakens as the distribution broadens in $v_\parallel$ (Figure 3d). Toward the overshoot ($z$: $80$-$110\,d_{\mathrm{e}}$), the parallel broadening becomes more pronounced (Figures~\ref{fig:vdf}e--f). In the downstream ($z$: $30$-$60\,d_{\mathrm{e}}$), the distribution develops a flat-top profile: the parallel extent is roughly twice the perpendicular extent, and the phase-space density is approximately uniform over $v_\parallel \in [-0.10c, 0.10c]$ (Figures 3g-i). We examine this progression from parallel broadening to a flat core in terms of nonuniform acceleration by the parallel electric field in Section~\ref{sec:flat_top}.

The relative beam density, defined as the beam density normalized to the local background electron density, increases toward the shock and reaches $\sim2\times 10^{-5}$ (Figures 3b-c). The beam speed remains close to 0.2c.

\section{Origin of the Whistler Turbulence}

Previous kinetic simulations have shown that the modified two-stream instability (MTSI) excites oblique electromagnetic whistler waves at quasi-perpendicular shocks with sufficiently large ion-to-electron mass ratios \citep{scholer2004,umeda2012}. Consistent with these results, we attribute the whistler branch identified in Case~S (Section~\ref{sec:turbulence}) to MTSI, with the cross-field drift of reflected ions relative to electrons providing the free energy.

The analyzed branch propagates at $30^\circ$ to the direction $z$ normal to the shock and at $\theta_{kB} \approx 85.67^\circ$ to the upstream magnetic field. Its larger normal than tangential wavevector component provides a substantial projection along the reflected-ion drift. At $t = 24000\,\omega_{\mathrm{pe}}^{-1}$, we estimate the characteristic reflected-ion velocity $\mathbf{u}_{\mathrm{r}}$ from the main body of the reflected-ion branch in the $z$--$v_{z}$ phase space across the shock transition region ($z\approx50$--$150\,d_{\mathrm{e}}$), using the laboratory-frame ion distributions. Its component along the shock normal is $u_{\mathrm{r},z} \approx 0.008c$. Assuming that the reflected-ion motion is predominantly along the shock normal, the component of $\mathbf{u}_{\mathrm{r}}$ perpendicular to $\mathbf{B}_0$ is $u_{\mathrm{r},\perp} \approx u_{\mathrm{r},z}\sin\theta_{Bn} \approx u_{\mathrm{r},z}$ at $\theta_{Bn} = 85^\circ$. Since the wave frequencies in Figure~2b are also evaluated in this frame, we use the approximate beam-resonance condition $\omega \approx \mathbf{k}\cdot\mathbf{u}_{\mathrm{r}}$. With $kd_{\mathrm{e}} \approx 0.4$ and $\mathbf{k}$ at about $30^\circ$ to the shock normal, this gives $\omega/\Omega_{\mathrm{ce}} \approx 0.028$, close to the peak frequencies of 0.03--0.04 in Figure~2b. This frequency agreement, together with the propagation geometry and whistler-mode properties, supports excitation by reflected-ion-driven MTSI.

\section{Trajectory Analysis of Electron Acceleration and VDF Formation}

We combine representative electron trajectories and statistical diagnostics in Case~S to explain the formation mechanisms of the upstream beam and downstream flat-top VDFs.

\subsection{Electron Acceleration at the Shock}

We randomly sampled one out of every 10000 electrons as tracers, yielding $\sim1.6 \times 10^{5}$ particles. From these trajectories, we selected three representative electrons at $t = 24000\,\omega_{\mathrm{pe}}^{-1}$: an upstream beam electron, an electron in the transition region, and a downstream flat-top electron (Figures 4a-c).

The upstream beam electron enters the ramp during $7000\text{--}12000\,\omega_{\mathrm{pe}}^{-1}$, when $v_\perp$ in the laboratory frame increases as the magnetic field strengthens. In the HT frame its pitch angle approaches $90^\circ$, and the electron is reflected upstream. As it moves into the weakening magnetic field, $v_\perp$ decreases while $v_\parallel$ increases to $\sim$0.22c, producing the field-aligned beam (Figure 4a). This evolution is consistent with shock-drift acceleration (SDA). The characteristic reflected-beam velocity scale in the upstream rest frame is $v_{\rm SDA}\approx2v_{\mathrm{sh}}/\cos\theta_{Bn}$ \citep{wu1984,amano2022}. With a shock speed relative to the upstream plasma $v_{\mathrm{sh}}\approx0.0115c$ and $\theta_{Bn}=85^\circ$, this gives $v_{\rm SDA}\approx0.26c$, comparable to the simulated beam speed. This estimate provides a characteristic scale rather than an exact prediction for the velocity increment of the selected electron.

\begin{figure}[htbp]
\centering\includegraphics[width=\linewidth,height=0.68\textheight,keepaspectratio]{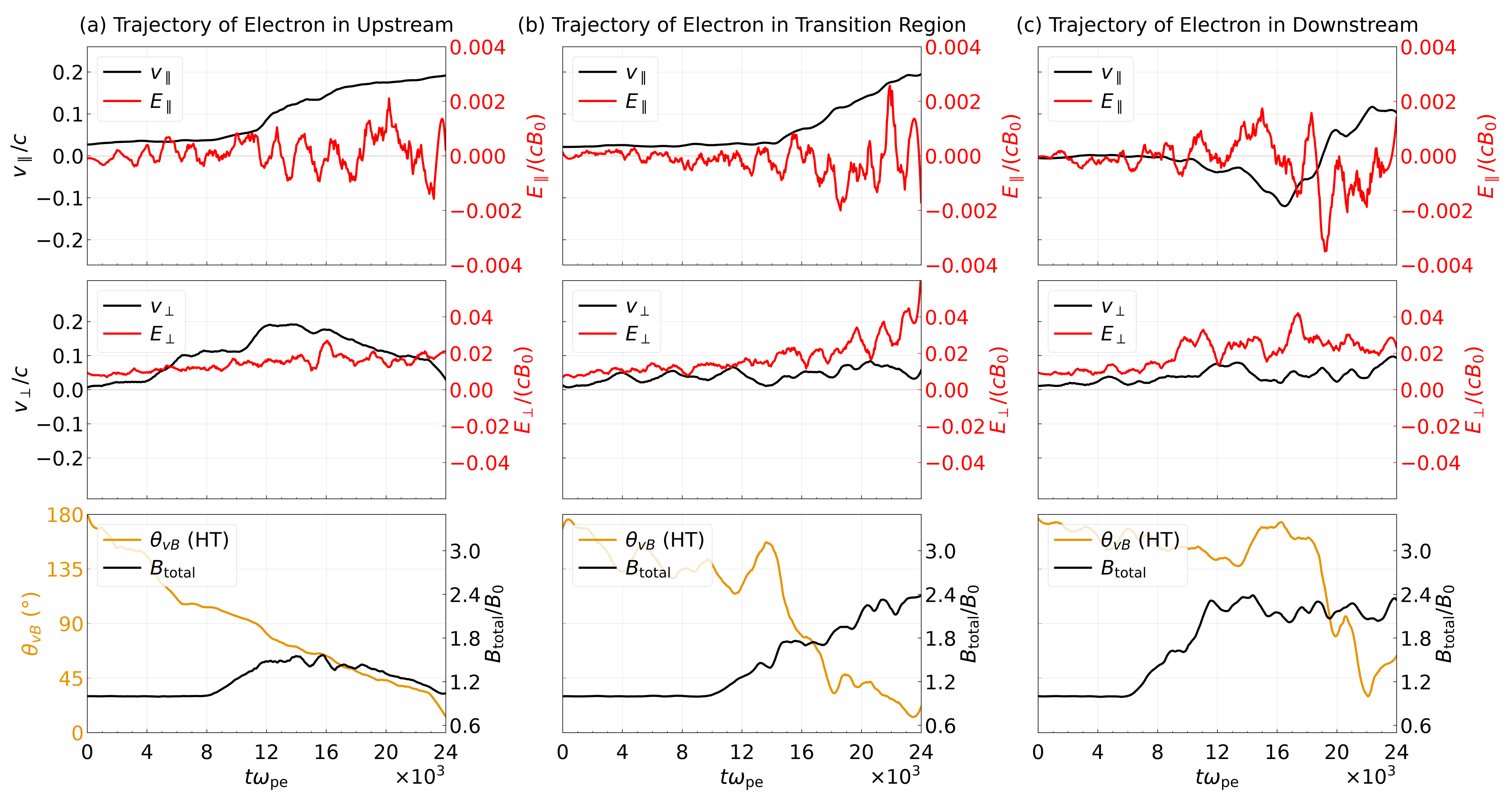}
\caption{Trajectories of electrons in Case~S selected at $t = 24000\,\omega_{\mathrm{pe}}^{-1}$: (a) upstream; (b) transition; (c) downstream. Top row: parallel velocity (black) and electric field (red); middle: magnitudes of the perpendicular velocity (black) and perpendicular electric field (red); bottom: pitch angle (yellow) and local magnetic field magnitude $B$ (black). Velocities and electric fields are normalized to $c$ and $cB_0$, respectively, and shown in the laboratory frame. Parallel and perpendicular components refer to the instantaneous local magnetic field. Pitch angles are measured in the de Hoffmann--Teller (HT) frame to identify signatures of magnetic mirroring; their evolution alone does not establish conservation of the magnetic moment or of the kinetic energy along the whole trajectory in the presence of cross-shock electric fields and electromagnetic fluctuations.}
\label{fig:trajectory}
\end{figure}

The electron selected in the transition region is reflected at the ramp during $10000\text{--}16000\,\omega_{\mathrm{pe}}^{-1}$ (Figure~\ref{fig:trajectory}b). However, strong electromagnetic fluctuations on electron-gyroradius scales cause the magnetic moment to become poorly conserved. During $16000\text{--}24000$ $\omega_{\mathrm{pe}}^{-1}$, $v_\perp$ of this electron remains nearly constant while $v_\parallel$ increases to $\sim$0.2c. Because electrons encounter the turbulent shock at different times and positions, they acquire different parallel velocity increments. These differences in acceleration histories are consistent with the parallel broadening observed in Figure 3d.

\subsection{Acceleration by the Parallel Electric Field and Flat-Top Formation}
\label{sec:flat_top}

The representative downstream electron encounters whistler turbulence after crossing the shock transition at $t \approx 12000\,\omega_{\mathrm{pe}}^{-1}$ (Figure~4c). Between $t = 14000$ and $16000\,\omega_{\mathrm{pe}}^{-1}$, a sustained parallel electric field $E_\parallel$ of a single polarity accelerates the electron from $-0.03c$ to $-0.10c$. Upon entering a region of opposite field polarity, the electron decelerates toward zero parallel velocity. From $t = 18000$ to $23000\,\omega_{\mathrm{pe}}^{-1}$, it crosses three to four such $E_\parallel$ patches, eventually reaching $v_\parallel \sim 0.1c$ with a net kinetic energy gain of $\sim2.6$ keV. This trajectory illustrates successive intervals of parallel acceleration and deceleration in electric field regions of different polarity. The accumulated parallel velocity change depends on the field history sampled along the trajectory.

\begin{figure}[htbp]
\centering\includegraphics[width=\linewidth]{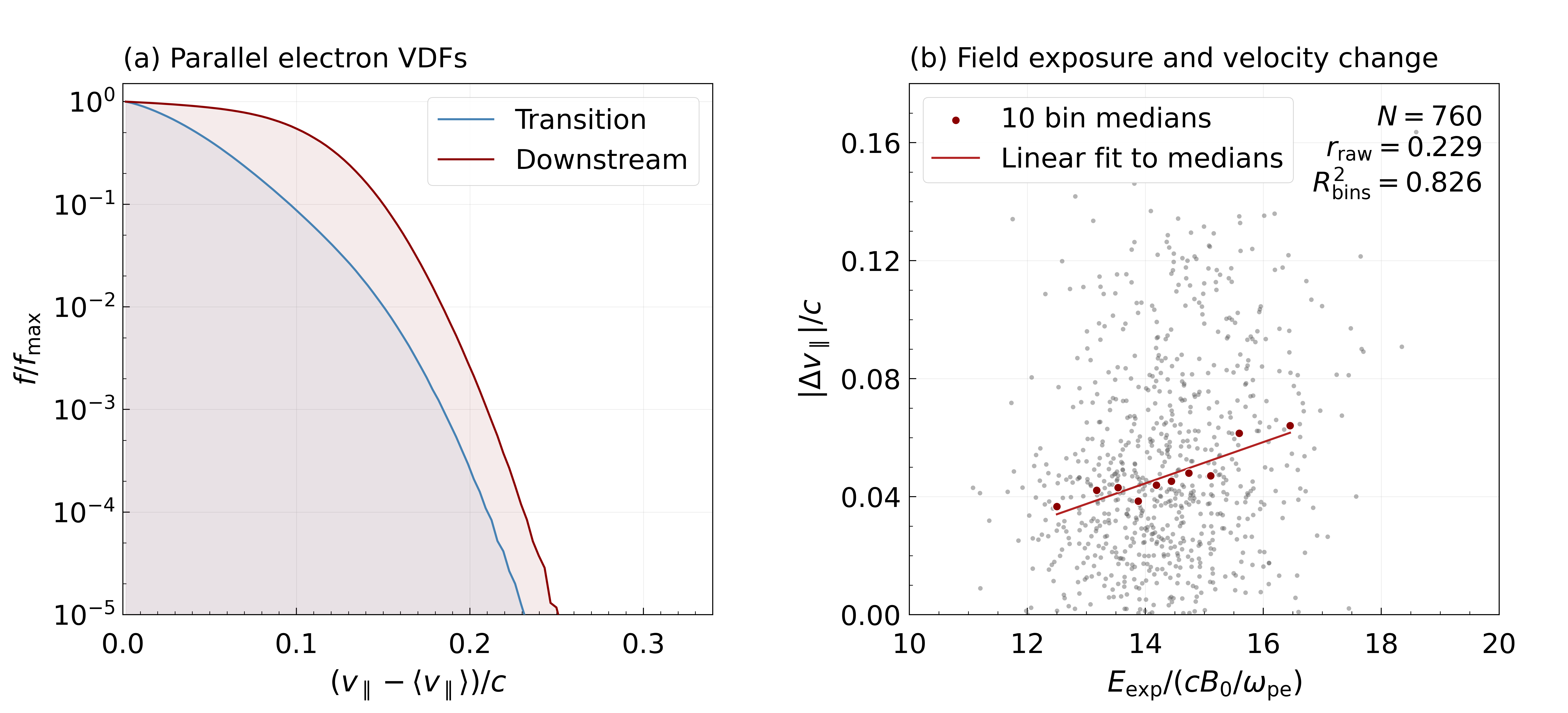}
\caption{Acceleration by the parallel electric field and downstream flat-top formation in Case~S. (a) Parallel VDFs in the transition ($t = 18000\,\omega_{\mathrm{pe}}^{-1}$, $z = 50$--$80\,d_{\mathrm{e}}$; blue) and downstream ($t = 24000\,\omega_{\mathrm{pe}}^{-1}$, $z = 10$--$40\,d_{\mathrm{e}}$; red), normalized to their peaks with the mean parallel velocity subtracted in each sampling region. The reduced distributions integrate over both perpendicular velocity components, and only $v_\parallel\geq0$ is shown. (b) Magnitude of the net parallel velocity change $|\Delta v_\parallel|$ and cumulative field exposure $E_{\rm exp} = \int |E_\parallel|\,dt$ over $[18000,24000]\,\omega_{\mathrm{pe}}^{-1}$ for 760 tracers. The selected electrons satisfy $z(18000) \geq 50\,d_{\mathrm{e}}$ and $z(24000) \in [10,50]\,d_{\mathrm{e}}$, with final laboratory-frame velocities $v_\parallel \in [0.05,0.3]c$ and $v_\perp \in [0.01,0.2]c$; particles reaching $z \leq 5\,d_{\mathrm{e}}$ at any recorded time are excluded. Panel (b) uses laboratory-frame velocities projected onto the instantaneous local magnetic field. Gray dots: individual electrons; red circles: medians in 10 equal-population $E_{\rm exp}$ bins; red line: least-squares fit to these medians. In the panel, $r_{\rm raw}$ is the Pearson coefficient over the individual electrons and $R^2_{\rm bins}$ is the coefficient of determination of the fit to the ten medians. The 95\% confidence interval of the slope is obtained from 10,000 bootstrap resamples of the individual electrons, each of which redefines the bins, recomputes the medians and refits both the slope and the intercept. Velocities are in units of $c$, and field exposure in $cB_0/\omega_{\mathrm{pe}}$. The absolute value in $|\Delta v_\parallel|$ is taken after the endpoint difference, whereas $E_{\rm exp}$ integrates the instantaneous field magnitude.}
\label{fig:velocityscattering}
\end{figure}

Figure~5a compares the parallel VDFs in the transition region ($t = 18000\,\omega_{\mathrm{pe}}^{-1}$, $z = 50$--$80\,d_{\mathrm{e}}$) and downstream ($t = 24000\,\omega_{\mathrm{pe}}^{-1}$, $z = 10$--$40\,d_{\mathrm{e}}$). The downstream distribution is broader and has a flatter core extending to $|v_\parallel| \sim 0.10c$. Figure~\ref{fig:velocityscattering}a displays only the $v_\parallel\geq0$ side of the reduced parallel VDF, with the mean parallel velocity subtracted in each sampling region. This distribution integrates over both perpendicular velocity components. These VDFs are sampled from separate spatial regions at the two times. We interpret the downstream flat-top as the outcome of nonuniform acceleration by the parallel electric field. Electrons following different trajectories encounter parallel electric fields with different strengths, polarities, and interaction durations. They therefore acquire different net changes in parallel velocity, and the downstream VDF is broader with a flatter core than the transition VDF (Figure~\ref{fig:velocityscattering}a).

To examine the statistical association with $E_\parallel$, Figure~5b compares the magnitude of the net parallel velocity change, $|\Delta v_\parallel| = |v_\parallel(t_1)-v_\parallel(t_0)|$, with the cumulative field exposure, $E_{\rm exp} = \int |E_\parallel|\,dt$, for 760 selected tracer electrons over $18000 \leq t\omega_{\mathrm{pe}} \leq 24000$. The absolute value of $\Delta v_\parallel$ is taken after the endpoint difference, so the cancellation accumulated along the trajectory is already reflected in it; only the sign of the net change is removed. In $E_{\rm exp}$, by contrast, the absolute value lies inside the integral, so fields of either polarity add up. $E_{\rm exp}$ therefore measures the field magnitude accumulated along a trajectory, not the net impulse, and it is not used here to quantify energy gain. A large $E_{\rm exp}$ does not imply a large net velocity change, because contributions of opposite polarity cancel in the signed field integral but not in $E_{\rm exp}$, and because the rotation of the local field direction adds a projection term to $\Delta v_\parallel$. The two quantities are weakly correlated (Pearson $r = 0.229$ for the individual (raw) electrons). A least-squares fit with a free intercept to the ten bin medians gives a slope of 0.00699 in the plotted units (95\% confidence interval 0.0038--0.0096). The intercept is retained in the fit but is not physically interpreted, because zero exposure lies outside the sampled range. The fit describes the typical change at a given exposure; the individual electrons are widely scattered. Taken together, the trajectory evidence, the broader and flatter downstream VDFs, and the weak positive association support nonuniform acceleration by the parallel electric field as the mechanism underlying downstream flat-top formation in Case~S.

\section{Conclusions and Discussion}

We used two-dimensional fully kinetic PIC simulations with the physical ion-to-electron mass ratio ($m_{\mathrm{i}}/m_{\mathrm{e}} = 1836$) to examine electron VDF formation at a low-$\beta$ quasi-perpendicular shock. Case~L encompasses at least one shock-reformation cycle. The field structure of Case~L at $t = 85000\,\omega_{\mathrm{pe}}^{-1}$ is consistent in overall morphology with that of Case~S at $t = 24000\,\omega_{\mathrm{pe}}^{-1}$, supporting the use of Case~S for the detailed electron-scale analysis.

In Case~S, the dispersion and polarization analyses identify highly oblique whistler-mode fluctuations in the shock transition. Their propagation geometry and approximate frequency agreement with the reflected-ion drift support excitation by MTSI. Representative electron trajectories support SDA as the origin of the upstream field-aligned beam. We attribute the downstream flat-top distributions in Case~S to nonuniform acceleration by the parallel electric field in the whistler-turbulent plasma. Individual electrons undergo successive intervals of acceleration and deceleration, with the accumulated parallel velocity change determined by the field strengths, polarities, and interaction durations sampled along their trajectories. These different acceleration histories redistribute electrons over parallel velocity, producing the broad, flattened core of the downstream VDF.

The importance of mass ratio for shock microinstabilities is established by earlier kinetic studies \citep{scholer2004,umeda2012}. The results of Case~S connect the electromagnetic wave environment at the physical mass ratio to the contrasting upstream and downstream electron distributions in the parameter regime considered here. Reduced-mass-ratio studies have emphasized ECDI-enhanced SDA \citep{guo2024} and acceleration by electron shear-flow structures \citep{guo2025}. The present results show that SDA and downstream parallel energization operate in a shock containing whistler turbulence. This comparison is consistent with known mass-ratio sensitivity, but isolating the effect of mass ratio for this setup requires otherwise matched simulations. Case~S therefore provides a complementary test of the electron dynamics under these shock conditions.

Cross-shock potential acceleration is a classical explanation for electron energization at shocks \citep{scudder1986,thomsen1987,scudder1995}. In Case~S, our analysis shows that its contribution to electron energization is smaller than that of turbulent $E_\parallel$. The SWE interpretation links reduced flat-top VDFs to projections of perpendicular ring or shell distributions \citep{stasiewicz2023}, whereas clear perpendicular ring-like signatures are not apparent in our VDFs. Electrostatic trapping also provides a route to distribution flattening \citep{oneil1965,zanelli2025}. The mechanism identified here centers on the different acceleration histories of electrons interacting with spatially and temporally varying parallel electric fields.

The conclusions apply to the simulated regime ($\beta_{\mathrm{e}}=\beta_{\mathrm{i}} = 0.01$, $M_{\mathrm{A}} \approx 5$, $\theta_{Bn} = 85^\circ$) and the analyzed stages of shock evolution. The tangential extent, $L_x = 90\,d_{\mathrm{e}} \approx 2.1\,d_{\mathrm{i}}$, restricts the accessible wavelengths of shock-surface structure, and the two-dimensional geometry restricts the available wavevectors. Case~L spans less than one upstream ion gyroperiod ($2\pi/\Omega_{\mathrm{ci}}$), so evolution over multiple ion gyroperiods remains untested. Longer-wavelength rippling, three-dimensional coupling, and electron energization over repeated reformation cycles also remain to be tested. Larger three-dimensional domains and controlled scans of mass ratio, $M_{\mathrm{A}}$, $\beta$, and $\theta_{Bn}$ will help establish how broadly the connection between whistler turbulence and electron VDF evolution applies.

\begin{acknowledgments}
This study is supported by the National Natural Science Foundation of China (NNSFC) grants (Nos. 42127804 and 42574218) and the National Key R\&D Program of China (No. 2025YFF0510701). The authors acknowledge the Beijing Super Cloud Computing Center (BSC-C, URL: \url{http://www.blsc.cn/}) for providing HPC resources, and the open-source Vector-PIC (VPIC) code provided by Los Alamos National Labs (LANL). We thank Lu Quanming for his constructive suggestions.
\end{acknowledgments}

\clearpage
\appendix
\restartappendixnumbering
\section{Long-term Shock Evolution and Downstream Turbulence}
\label{app:diagnostics}

\subsection{Late-time Shock Structure in Case L}

\begin{figure}[htbp]
\centering
\includegraphics[width=\hsize]{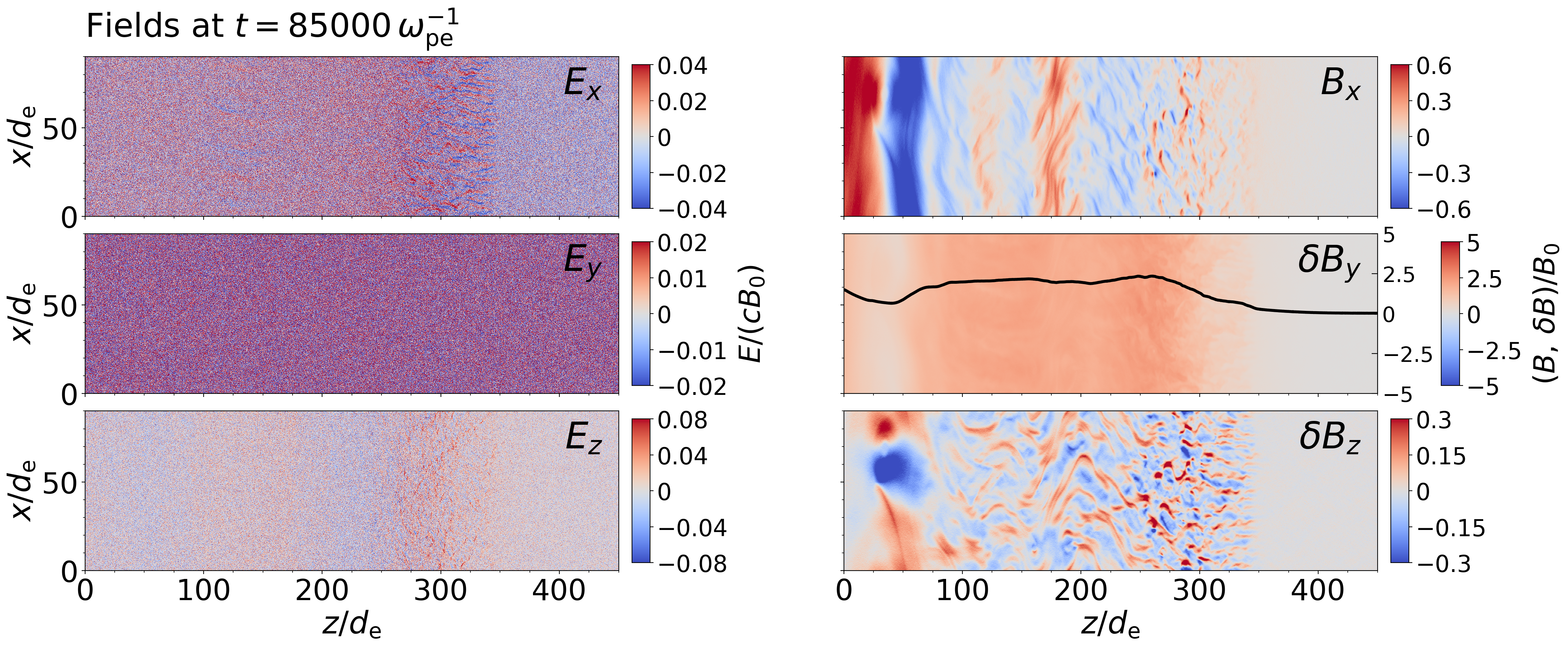}
\caption{Electromagnetic field structure at $t=85000\,\omega_{\mathrm{pe}}^{-1}$ in Case~L. The displayed region is $0\leq z\leq450\,d_{\mathrm{e}}$. The left column shows $E_x$, $E_y$, and $E_z$, and the right column shows $B_x$, $\delta B_y$, and $\delta B_z$, with the initial magnetic field subtracted in the latter two panels. Electric and magnetic fields are normalized to $cB_0$ and $B_0$, respectively. The black curve shows $\langle\delta B_y\rangle_x/B_0$ as a function of $z$, where $\delta B_y=B_y(t)-B_y(0)$ and the average is over the transverse direction $x$. Its amplitude is read from the independent right-hand vertical axis; the color map uses the spatial coordinates $z/d_{\mathrm{e}}$ and $x/d_{\mathrm{e}}$. Color limits and spatial ranges differ from those in Figure~\ref{fig:fields}b. The corresponding animation (\texttt{movieS1.mp4}; playback duration 20.2 s) covers $t=0$ to $100000\,\omega_{\mathrm{pe}}^{-1}$ and is available through Science Data Bank (DOI: \href{https://doi.org/10.57760/sciencedb.016yt}{10.57760/sciencedb.016yt}).}
\label{fig:shockevolution}
\end{figure}

Figure~\ref{fig:shockevolution} shows the electromagnetic fields at $t=85000\,\omega_{\mathrm{pe}}^{-1}$ in Case~L. This time is marked by the upper white dashed line in Figure~\ref{fig:fields}a. The snapshot complements the time history by showing the two-dimensional structure of the magnetic compression and the associated field fluctuations at a later stage. It is compared with the snapshot from Case~S in Figure~\ref{fig:fields}b in Section~\ref{sec:shock_structure}. The corresponding animation (\texttt{movieS1.mp4}) shows the two-dimensional electromagnetic field evolution in Case~L from $t=0$ to $100000\,\omega_{\mathrm{pe}}^{-1}$, providing the temporal context for the snapshot in Figure~\ref{fig:shockevolution}. The video is available through Science Data Bank (DOI: \href{https://doi.org/10.57760/sciencedb.016yt}{10.57760/sciencedb.016yt}).

\clearpage
\subsection{Downstream Turbulence in Case S}

\begin{figure}[htbp]
\centering\includegraphics[width=\linewidth,height=0.68\textheight,keepaspectratio]{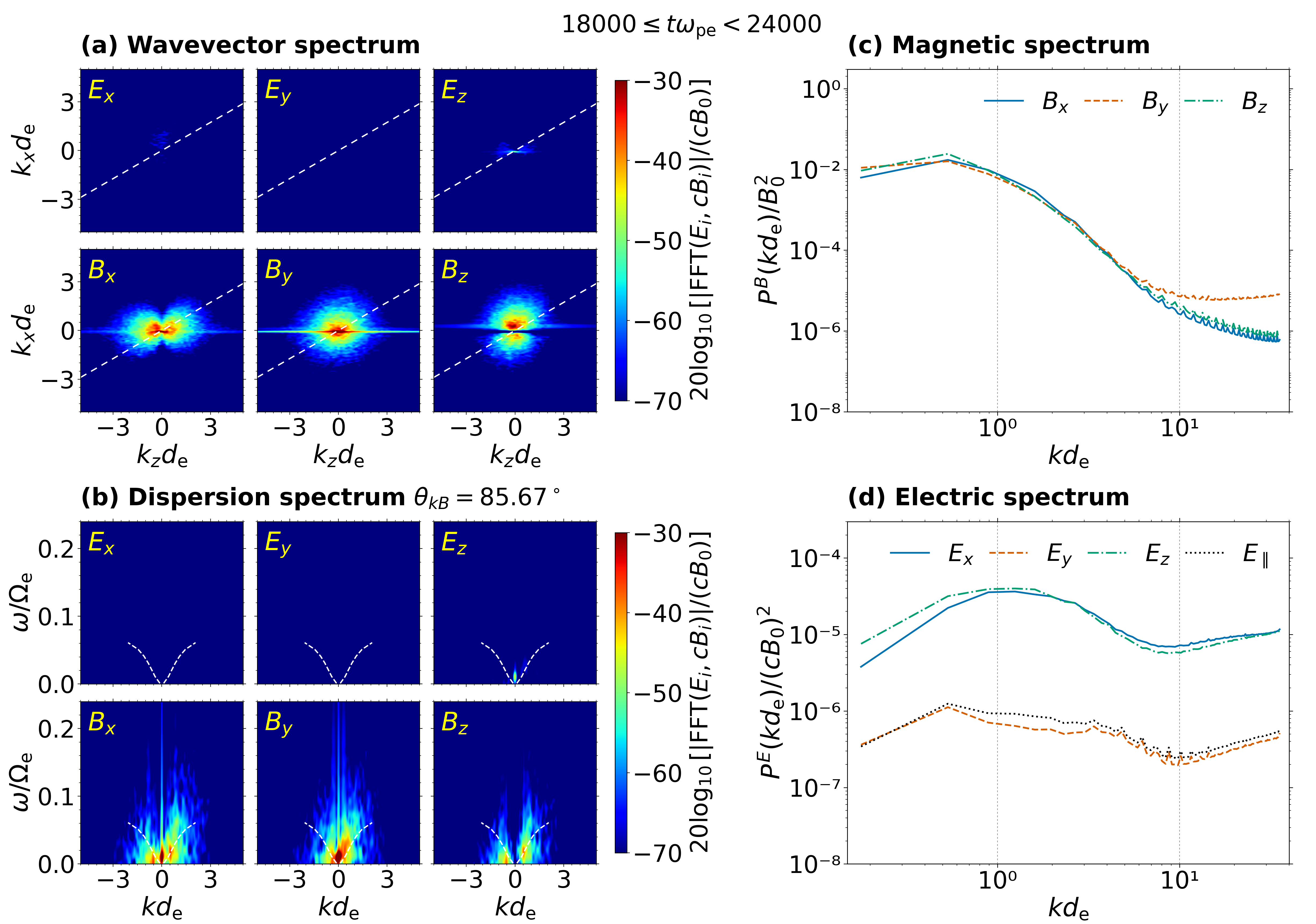}
\caption{Wave modes and spectra in the downstream region of Case~S ($18000 \leq t\omega_{\mathrm{pe}} < 24000$; $x$: $0$--$90\,d_{\mathrm{e}}$, $z$: $10\text{--}50\,d_{\mathrm{e}}$). The panel layout is identical to Figure~\ref{fig:rampturbulence}. (a) Frequency-maximized Fourier amplitudes in $k_x$--$k_z$ space, shown on the logarithmic color scale. White dashed lines mark the $30^\circ$ slice measured from the positive $k_z$ axis. (b) $\omega$--$k$ spectrum along this slice, corresponding to $\theta_{kB}=85.67^\circ$; white dashed curves show the cold-plasma whistler dispersion. Wavenumbers are expressed as $kd_{\mathrm{e}}$, and frequency as $\omega/\Omega_{\mathrm{e}}$, with $\Omega_{\mathrm{e}}=\Omega_{\mathrm{ce}}$. (c) Magnetic spectra of $B_x$, $B_y$, and $B_z$. (d) Electric spectra of $E_x$, $E_y$, and $E_z$, with $E_\parallel$ relative to $B_0$ shown as a black dotted curve. Panels (c)--(d) show radial shell-integrated power densities per unit $kd_{\mathrm{e}}$, normalized to $B_0^2$ and $(cB_0)^2$, respectively; both signs of nonzero temporal frequency are included, while the zero-frequency component is excluded. Gray dashed vertical lines mark $1 \leq kd_{\mathrm{e}} \leq 10$. Fits of $P(kd_{\mathrm{e}})\propto(kd_{\mathrm{e}})^{-\alpha}$ over this interval give $\alpha_{Bx} = 3.90$, $\alpha_{By} = 3.25$, $\alpha_{Bz} = 3.57$, $\alpha_{Ex} = 0.98$, $\alpha_{Ey} = 0.61$, $\alpha_{Ez} = 1.14$, $\alpha_{E\parallel} = 0.72$.}
\label{fig:downstreamturbulence}
\end{figure}

Figure~\ref{fig:downstreamturbulence} presents the downstream wave-mode and spectral analysis from Case~S over $18000\leq t\omega_{\mathrm{pe}}<24000$, complementing the analysis of the transition region in Figure~\ref{fig:rampturbulence}. The downstream turbulence is dominated by whistler modes, with wave spectral characteristics qualitatively similar to those in the transition region (Figure~\ref{fig:rampturbulence}).

\clearpage

\bibliography{refs}{}
\bibliographystyle{aasjournalv7.1}

\end{document}